\documentclass[pre,twocolumn,showpacs,floatfix,amsmath,amssymb]{revtex4}
\usepackage{graphicx}
\usepackage{dcolumn}
\usepackage{color}
\usepackage{bm}
\usepackage{verbatim}
\usepackage{multirow}
\usepackage{amsmath,amssymb,graphicx}
\usepackage{epsfig}
\usepackage{enumerate}
\usepackage{array}
\usepackage{multirow}
\begin{document}
\title {Control of quantum thermodynamic behaviour of a charged magneto oscillator with momentum dissipation}
\author{Asam Rajesh and Malay Bandyopadhyay}
\affiliation{School of Basic Sciences, Indian Institue of Technology Bhubaneswar, Bhubaneswar, India 751007}

\vskip-2.8cm
\date{\today}
\vskip-0.9cm

\begin{abstract}
\vskip 0.5cm
In this work, we expose the role of environment, confinement and external magnetic field ($B$) in determining the low temperature thermodynamic behaviour in the context of cyclotron motion of a charged oscillator with anomalous dissipative coupling involving the momentum instead of the much studied coordinate coupling. Explicit expressions for different quantum thermodynamic functions (QTF) are obtained at low temperatures for different quantum heat bath characterized by spectral density function, $\mu(\omega)$. The power law fall of different QTF are in conformity with third law of thermodynamics. But, the sensitiveness of decay i.e. the power of the power law decay explicitly depends on $\mu(\omega)$. We also separately discuss the influence of confinement and magnetic field on the low temperature behavior of different QTF. In this process we demonstrate how to control low temperature behaviour of  anomalous dissipative quantum systems  by varying confining length $a$, $B$ and the temperature $T$. Momentum dissipation reduces effective mass of the system and we also discuss its effect on different QTF at low temperatures.
\end{abstract}
\maketitle
\section{Introduction}
A most common and effective approach to deal quantum dissipation is the system-plus-reservoir (SPR) model \cite{a,b,c,d} where dissipation arises from the linear coupling of some system's variable with a reservoir (or environment or bath) which is modelled as a collection of independent quantum harmonic oscillators which are individually weakly perturbed by the system so that the reservoir can be assumed to be at equilibrium. Usually the system's variable is taken to be a function of system's coordinate. The model can indeed be described by Langevin equation \cite{e} after eliminating bath variables and a connection between bath spectrum and the dissipative memory function can be established. This connection between phenomenological description and microscopic details of the bath is very much valuable as it gives us the connectivity between quantum thermodynamic behaviour of the system and phenomenology.\\
\indent
The standard SPR model for quantum dissipation can meaningfully be generalized to the complementary possibility of coupling of a quantum system to a quantum mechanical heat bath through the momentum variables. It is now appropriate to discuss about previous studies on quantum dissipation where coupling occurs through momentum variables. In an early paper, Leggett \cite{f} discussed about normal dissipation and anomalous dissipation by considering two possible coupling terms ${\bf{q}}\sum_jd_j{\bf{p_j}}+{\bf{p}}\sum_jc_j{\bf{q_j}}$ where $\bf{q}$, $\bf{p}$ are system coordinate and momentum variables respectively and ${\bf{q_j}}$ , ${\bf{p_j}}$ , $d_j$ and $c_j$ are coordinate, momentum and coupling constants of the $j^{th}$ bath oscillator respectively. But, our Hamiltonian as given in Eq.(1 ) has no resemblance with any of the two coupling terms considered by Leggett. Our model Hamiltonian (1) has the coupling term ${\bf{p}}\sum_j g_j {\bf{p_j}}$ which definitely leads to different dynamics than that of Leggett. Recently, Cuccoli et al\cite{g} and Ankerhold et al \cite{h} studied momentum coupling model which has some resemblance to our model. Infact, our model Hamiltonian can exactly be cast  into the model Hamiltonian considered by Ankerhold et al\cite{h} by an Unitary transformation. As mentioned earlier the dissipation arises from this type of model is completely different from that obtained from the model studied by Leggett. To differentiate between the two models, we follow the terminology of Ankerhold et al \cite{h} and use the term momentum dissipation rather than anomalous dissipation used by Leggett. Besides some fundamental theoretical issues, the discussion about this momentum dissipation with the model Hamiltonian (1) has some physical relevance. Makhnovskii and Pollok \cite{makhno} showed that momentum dissipation leads to stochastic acceleration \cite{fermi,sturrock} without violating second law of thermodynamics \cite{cole}. In general, momentum dissipation reduces effective mass of the system and leads to anti-intuitive results of amplifying quantum effect rather than destroy it. We have discussed  the effect of reduction of effective mass on different QTF in the context of charged dissipative magneto-oscillator. Although, there are several instances for which mass renormalization can be negative for the position-position coupling \cite{ingold1,ingold2}.  F. Sols et al \cite{sols1,sols2} have discussed about momentum dissipation and local gauge invariance in the context of spin 1/2 impurity in an antiferromagnetic environment. The gauge invariance of the Hamiltonian is taken care in our case also.\\
\indent
Recently, we derive the quantum Langevin equation (QLE) satisfied by the particle coordinate operator for a charged quantum oscillator moving in a harmonic potential ($\omega_0$) with an external uniform magnetic field ($B$) and in the presence of momentum dissipation (the Hamiltonian (1)) and the same system is considered in this work \cite{j}. Later, we calculate explicitly the equilibrium free energy for the same Hamiltonian (1) in Ref. \cite{k}. In this work, we extend our previous studies \cite{j,k} by incorporating the discussion of low temperature behaviour of different QTF for different realistic situations and the control of low temperature behaviour of different QTF by varying external parameters. Different control parameters like magnetic field($B$) and confining length($a$) are identified. Low temperature behaviour of different QTF are analyzed for different realistic situations which are captured through different realistic bath spectrum $\mu(\omega)$. The effect of reduction of effective mass of the system on different QTF is also analyzed.\\
\indent
It is shown in Ref. \cite{k} that the equilibrium free energy for the Hamiltonian (1) can be expressed as $F_0(T,B)= F_0(T,0)+ \Delta_1F_0(T,B)+\Delta_2F_0(T,B)$, where $T$ and $B$ are temperature and external magnetic field of the system respectively. In this paper, we explicitly show how to control the low temperature behaviour of different QTF obtainable from $F_0(T,B)$ by varying three characteristic frequencies of the system : thermal frequency $\omega_{th}=\frac{k_BT}{\hbar}$, cyclotron frequency $\omega_c=\frac{qB}{m}$, and confining potential frequency $\omega_0=\frac{\pi^2\hbar}{2ma^2}$. Here $k_B$, $\hbar$, $c$ are, respectively, the Boltzman constant, Planck constant and the velocity of light in free space, while $q$, $m$ and $a$ are charge, mass, and confining length of the particle respectively.  It is now useful to state our principal results at the outset : Considering a generalized quantum heat bath, we find three different regimes depending on the ratio $\frac{\omega_c\omega_{th}}{\omega_0^2}$ of three characteristic frequencies $\omega_c$, $\omega_0$ and $\omega_{th}$ of the system : (i) Region I where $F_0(T,0)$ dominates the low temperature thermodynamic properties; (ii)Region II, a fuzzy regime, where all the three terms of free energy are important in determining low temperature thermodynamic properties; (iii) Region III where the power of the power law fall of different QTF are determined by $\Delta_1F_0(T,B)=\Delta_2F_0(T,B)$; (iv)In the absence of confining potential ($\omega_0=0$) the low temperature behaviour is fully determined by $F_0(T,0)$; (v) Unlike the normal dissipation, the low temperature behaviour for $\omega_0=0$ is independent of $\omega_c$ with momentum dissipation for the arbitrary heat bath; (vi)Finally,  momentum dissipation decreases mass of the system (anti-intuitive quantum effects) and as a result the quantum contribution to different QTF increases with the increase of dissipation. Although, there are certainly a number of studies which demonstrate that the quantum corrections arise for a free particle only in the presence of position-position coupling \cite{ingold3,ingold4,ingold5}. On the other hand, all the above mentioned results hold for the radiation heat bath, but the low temperatures phase diagram is fully controlled by the ratio $\frac{\omega_c}{\omega_0}$. Like the normal dissipation, the low temperature behaviour of different QTF for the radiation bath depends on $\omega_c$ and friction constant $\gamma$ for the case $\omega_0=0$ with momentum dissipation.\\
\indent
With this we summarize the construction of our paper. In the following section, we describe our model and summarize the basic expressions which are required for our further calculations. In Sec. III, we consider a generalized quantum heat bath spectrum and derive several QTF at low temperatures for our model system. As a matter of fact we identify different external control parameters like  $B$ and $a$ to control the low temperature behaviour of our model system. Separately we discuss the case for $\omega_0=0$. Further, we consider low temperature behaviour for the charged magneto-oscillator coupled with radiation bath through momentum variables in this Sec. (III).  We also discuss the effect of the negative renormalization of the mass of the system due to momentum coupling and its effect on different QTF at low temperatures in this section. We conclude this section with a possible demonstration of our control mechanism of different QTF for a physically realizable system. The paper is concluded in Sec. (IV).
\section{Model System \& Basic Expressions}
We consider the dissipative dynamics of a charged quantum oscillator in the presence of an external magnetic field. Such a situation is often arises in many problems of theoretical and experimental relevance,e.g., Landau dissipative diamagnetism \cite{l}, quantum Hall effect \cite{m}, and high temperature superconductivity \cite{n}. Recently, we analyze the dissipative dynamics of such system by considering bilinear coupling  between the system and the environment through momentum variables by invoking a gauge-invariant SPR model \cite{j,k}. The Hamiltonian of the system is
\begin{eqnarray}
&&H_0=\frac{1}{2m}\Big({\bf{p}}-\frac{q}{c}{\bf{A}}\Big)^2+\frac{1}{2}m\omega_0^2{\bf{r}}^2\nonumber \\
&&+\sum_{j=1}^N\Big\lbrack\frac{1}{2m_j}\Big({\bf{p_j}}-g_j{\bf{p}}+\frac{g_jq}{c}{\bf{A}}\Big)^2+\frac{1}{2}m_j\omega_j^2{\bf{q_j}}^2\Big\rbrack,
\end{eqnarray}
where q,m, ${\bf{p}}$,${\bf{r}}$ are, respectively, the charge, the mass,the momentum operator and the coordinate operator of the particle, while $\omega_0$ is the frequency characterizing its motion in the harmonic potential. The $j$th heat-bath oscillator has mass $m_j$, frequency
$\omega_j$, coordinate operator ${\bf q}_j$, and momentum operator ${\bf
p}_j$. The dimensionless parameter $g_j$ describes the coupling between
the particle and the $j$th oscillator. The speed of light in vacuum is
denoted by $c$. The vector potential ${\bf A}={\bf A}({\bf r})$ is
related to the uniform external magnetic field ${\bf B}=(B_x,B_y,B_z)$ through ${\bf
B}=\nabla \times {\bf A}({\bf r})$. The field has the magnitude
$B=\sqrt{B_x^2+B_y^2+B_z^2}$.
The commutation relations for the different coordinate and momentum operators are
\begin{equation}
[r_\alpha,p_\beta]=i\hbar \delta_{\alpha \beta}, [q_{j\alpha},p_{k\beta}]=i\hbar \delta_{jk}\delta_{\alpha \beta},
\end{equation}
while all other commutators vanish. In the above equation, $\delta_{jk}$
denotes the Kronecker Delta function. Here, Greek indices ($\alpha, \beta, \ldots$) refer to the three spatial directions, while Roman indices ($i,j,k,\ldots$) represent the heat-bath oscillators. Let us remark that momentum-momentum coupling has been considered earlier in the literature
\cite{f}, and, in particular, to model the physical situation
of a single Josephson junction interacting with the blackbody
electromagnetic field in the dipole approximation
\cite{g,h}. Our model Hamiltonian is similar to
that considered in Refs. \cite{g,h}, the additional interesting feature
that we consider here is the inclusion of the effects of an external magnetic field.
In Ref. \cite{k}, we derived the equilibrium free energy of Hamiltonian (1) in the following form :
\begin{equation}
F_0(T,B)=F_0(T,0)+\Delta_1 F_0(T,B)+\Delta_2 F_0(T,B),
\end{equation}
where
\begin{eqnarray}
F_0(T,0)=\frac{3}{\pi}\int_0^\infty d\omega~ f(\omega,T){\rm
Im}\Big[\frac{d}{d\omega}\ln \alpha^{(0)}(\omega)\Big]
\end{eqnarray}
is the free energy of the charged particle in the absence of the
magnetic field. The contribution from the latter is contained in the two
terms $\Delta_1 F_0(T,B)$ and $\Delta_2 F_0(T,B)$, given by
\begin{eqnarray}
&&\Delta_1 F_0(T,B)=-\frac{1}{\pi}\int_0^\infty d\omega~ f(\omega,T){\rm
Im}\Big[\frac{d}{d\omega}\ln \Big\{1-\nonumber \\
&&(G(\omega))^2\Big(\frac{\omega B
q}{c}\Big)^2[\alpha^{(0)}(\omega)]^2\Big\}\Big],
\end{eqnarray}
and
\begin{eqnarray}
&&\Delta_2 F_0(T,B)=\frac{1}{\pi}\int_0^\infty d\omega~ f(\omega,T){\rm Im}\Big[[\alpha^{(0)}(\omega)]^2\Big(\frac{\omega
qB}{c}\Big)^2\nonumber \\
&\times&\Big(\frac{d(G(\omega))^2}{d\omega}\Big)\Big\{1-\Big(\frac{\omega B
qG(\omega)}{c}\Big)^2[\alpha^{(0)}(\omega)]^2\Big\}^{-1}\Big].
\end{eqnarray}
Here $f(\omega,T)$ is the free energy of a free oscillator of frequency
$\omega$:
\begin{equation}
f(\omega,T)=k_BT \ln \Big[2 \sinh\Big[\frac{\hbar\omega}{2k_BT}\Big]\Big],
\end{equation}
$\alpha^{(0)}(\omega)=[-m_r\omega^2+m\omega_0^2G(\omega)]^{-1}$ is the susceptibility in the absence of the magnetic field, $G(\omega)=1-\sum_{j=1}^N\frac{(g_j)^2m_r\omega_j^2}{m_j(\omega_j^2-\omega^2)}$ and renormalized mass $m_r=\frac{m}{1+\sum_{j=1}^N\frac{g_j^2m}{m_j}}$.
Let us now comment on the form of the free energy(3) with respect to that for
coordinate-coordinate coupling between the particle and the heat-bath
oscillators, obtained in Ref. \cite{li}. In the latter case, the free
energy is given by
\begin{equation}
F_0(T,B)=F_0(T,0)+\Delta_1 F_0(T,B),
\end{equation}
where $F_0(T,0)$ has the same form as in Eq.(4),
while $\Delta_1 F_0(T,B)$ is of the same form as given in Eq. (5) except the $G(\omega)$ term. This is to be remembered that the existence of $G(\omega)$ and negative renormalized mass $m_r$ in the free energy for momentum dissipation change thermodynamic properties dramatically. One can also compare our results Eqs. (3)-(6) with that of Wang {\em et. al} \cite{wang} who have discussed the momentum dissipation in the context of a dissipative oscillator. One can easily show that Eqs. (5) and (6) identically vanish for vanishing magnetic field $B$.Thus, our results exactly matches with that of Eq. (4) of Wang {\em et. al} \cite{wang} except a prefactor of 3 which is coming due to 3-dimensions considered in our problem.
\section{Different QTF with different illustrative $\mu(\omega)$}
In this section, we utilize the main results  obtained in Ref. \cite{k} i.e. the equilibrium free energy for Hamiltonian (1) to obtain and analyze different QTF at low temperatures for the dissipative cyclotron motion with momentum dissipation. To proceed, we need to express $G(\omega)$ and $\alpha^{0}(\omega)$ in terms of the Fourier transform of the diagonal part of the memory function, given by \cite{j} $\mu_d(\omega)=i\sum_{j=1}^N\frac{g_j^2mm_r\omega\omega_0^2}{m_j(\omega^2-\omega_j^2)}$. Thus we can have the free oscillator susceptibility
\begin{equation}
\alpha^{(0)}(\omega)=\frac{1}{m_r(\omega_0^2-\omega^2)-i\omega\mu_d(\omega)},
\end{equation}
and
\begin{equation}
G(\omega)=\frac{m_r}{m}-\frac{i\omega\mu_d(\omega)}{m\omega_0^2}.
\end{equation}
At this point, we consider different $\mu_d(\omega)$ for different heat bath spectrum and derive explicit expression of different QTF at low temperatures.
\subsection{Arbitrary heat bath}
We consider here a very general class of heat bath for which in the small $\omega$ regime we have $\mu_d(\omega)\simeq mb^{1-\nu}\omega^{\nu}$, with b is a positive constant with the dimension of frequency \cite{rmp}. The Ohmic, sub-Ohmic and super-Ohmic heat bath spectra are characterized by considering $\nu=1$, $0<\nu<1$, and $\nu>1$, respectively. These three cases are also relevant for a real physical system. To describe quantum tunneling in a metallic environment one can use the Ohmic spectrum \cite{b}. The super-Ohmic spectrum corresponds to the phonon bath in $d>1$ spatial dimensions and it refers to $\nu=d$ or $\nu=d+ 2$ cases depending on the underlying symmetry of the strain field \cite{b}. The sub-Ohmic spectrum is useful in describing the type of noise in some solid state devices or 1/f noise in Josephson junction \cite{o}. This is to mention here that there is no harm in choosing Ohmic, sub-Ohmic or super-Ohmic bath in the context of momentum dissipation until and unless these choices make any conflict with the conditions (49) and (50) of the heat bath spectrum obtained in Ref. \cite{j}. We can rearrange the free energy expressions as follows :
\begin{eqnarray}
F_0(T,B)=F_0(T,0)+\Delta_1 F_0(T,B)+\Delta_2 F_0(T,B)\nonumber \\
=\frac{1}{\pi}\int_0^{\infty}d\omega f(\omega,T)[3I_0-I_1+I_2],
\end{eqnarray}
with
\begin{eqnarray}
I_0&=&\Im\Big[\frac{d}{d\omega}\ln\alpha^{(0)}(\omega)\Big], \nonumber \\
I_1&=&\Im\Big[\frac{d}{d\omega}\ln\Big\lbrace 1-(G(\omega))^2\Big(\frac{qB\omega}{c}\Big)^2[\alpha^{(0)}(\omega)]^2\Big\rbrace\Big\rbrack, \nonumber \\ I_2&=&\Im\Big[\frac{[\alpha^{(0)}(\omega)]^2\Big(\frac{\omega qB}{c}\Big)^2\Big(\frac{d(G(\omega))^2}{d\omega}\Big)}{\Big\lbrace1-(G(\omega))^2\Big(\frac{qB\omega}{c}\Big)^2[\alpha^{(0)}(\omega)]^2\Big\rbrace}\Big].
\end{eqnarray}
Now, $f(\omega,T)$ vanishes exponentially for $\omega>>\frac{k_BT}{\hbar}$. Therefore, in order to evaluate the free energy of the dissipative charged oscillator at low temperatures, we need to consider only low-$\omega$ contributions of integrands in evaluating the integral in Eq. (11). With this we can show that at low frequencies the magnetic field independent integrand $I_0$ becomes :
\begin{equation}
\lim_{\omega\rightarrow 0}I_0(\omega)\simeq \frac{C(1+\nu)}{\omega_0^2}\omega^{\nu},
\end{equation}
with $C=\frac{m}{m_r}b^{1-\nu}\cos\Big(\frac{\nu\pi}{2}\Big)$. On the other hand, we obtain for the  magnetic field dependent integrands $I_1$ and $I_2$ as follows  :
\begin{eqnarray}
\lim_{\omega\rightarrow 0}I_1= \frac{2C(\nu+1)\omega_c^2}{\omega_0^6}\omega^{\nu+2}
=\lim_{\omega\rightarrow 0}I_2
\end{eqnarray}
Now, we use the result
\begin{equation}
\int_{0}^{\infty}dy y^{\nu}\log(1-e^{-y})=-\Gamma(\nu+1)\zeta(\nu+2),
\end{equation}
where $\Gamma(z)$ is the gamma function, while $\zeta(z)$ is the Riemann Zeta function, to obtain the free energy at low temperatures :
\begin{widetext}
\begin{eqnarray}
F_0(T,0)\simeq -\frac{3\Gamma(\nu+2)\zeta(\nu+2)\cos\Big(\frac{\nu\pi}{2}\Big)m\hbar b}{m_r\pi}\Big(\frac{b}{\omega_0}\Big)^2\Big(\frac{\omega_{th}}{b}\Big)^{\nu+2}\nonumber \\
\Delta_1F_0(T,B)\simeq -\frac{2(\nu+1)\Gamma(\nu+3)\zeta(\nu+4)\cos\Big(\frac{\nu\pi}{2}\Big)m\hbar b}{m_r\pi}\Big(\frac{\omega_c}{\omega_0}\Big)^2\Big(\frac{b}{\omega_0}\Big)^4\Big(\frac{\omega_{th}}{b}\Big)^{\nu+4}
=\Delta_2F_0(T,B).
\end{eqnarray}
\end{widetext}
We have already discussed that one can identify three characteristic frequencies of the system : (i)thermal frequency $\omega_{th}$, (ii) cyclotron frequency $\omega_c$, and (iii) the confining potential frequency $\omega_0$. The ratio of $\Delta_1F_0(T,B)$ or $\Delta_2F_0(T,B)$ with respect to $F_0(T,0)$ can be expressed in terms of the above mentioned three frequencies :
\begin{equation}
\frac{\Delta_1F_0(T,B)}{F_0(T,0)}= \frac{\Delta_2F_0(T,B)}{F_0(T,0)}\simeq \Big(\frac{\omega_c}{\omega_0}\Big)^2\Big(\frac{\omega_{th}}{\omega_0}\Big)^2.
\end{equation}
Thus, one can control the ratios of different terms of free energies and henceforth the thermodynamic behaviour by varying the external parameters like temperature (T) associated with thermal frequency, magnetic field (B) for cyclotron frequency and confining well length (a) related to confining potential frequency. One can roughly identify three different regimes depending on the ratio $\frac{\omega_{th}\omega_c}{\omega_0^2}$. Here, we must admit that it is impossible to identify a well defined boundary between two different regimes for the present case. But,one can set a boundary between two regimes if the ratio in Eq. (17) differs by an order of magnitude. we have tried to distinguish three regimes based on the dependence of three terms of the free energy (Eq 11) on the ratios $\frac{\omega_c}{\omega_0}$ and $\frac{\omega_{th}}{\omega_0}$. As we have already discussed that all our discussion is valid at very low temperatures, so $\omega_{th}^{max}\sim 10^5 Hz$ with $T_{max}\sim 10^{-6}K$. Similarly, as we are interested to confine our charged particle utmost inside a microstructure (length $a_{max}\sim 10^{-6}$) which leads to $\omega_0=\frac{\hbar\pi^2}{2ma^2}\sim 10^6 Hz$. Thus, one can set upper limit of the ratio $\frac{\omega_{th}}{\omega_0}\sim 0.1$. Similarly, the lower limit of this ratio for confinement of the charged particle inside a nanostructure (dimension $a\sim 10^{-9}m$) at temperatures $T_{min}\sim 10^{-9}K$ can be set at $\frac{\omega_{th}}{\omega_0}\sim 10^{-10}$. On the other hand, the upper limit of the ratio $\frac{\omega_c}{\omega_0} \sim 10^2$ for a maximum magnetic field $B_{max}\sim 10 T$ and confinement length $a_{max}\sim 10^{-6}m$. The lower limit for $\frac{\omega_c}{\omega_0}$ is set to $\sim 10^{-8}$ for $B_{min}\sim 10^{-3}T$ and confinement length $a_min\sim 10^{-9}m$ (nanostructure). After setting the upper and lower bound of the ratios $\frac{\omega_{th}}{\omega_0}$ and $\frac{\omega_c}{\omega_0}$, one can obtain three different regimes by varying $B$ and $a$ . For this purpose our crucial equation is Eq. (17). We have already shown that at low temperatures $\Delta_1F_0(T,B) = \Delta_2F_0(T,B)$ for the arbitrary heat bath spectrum. From Eq. (17), we observe that if $\frac{\omega_c}{\omega_0}\frac{\omega_{th}}{\omega_0}<<1$, $F_0(T,0)>>\Delta_1F_0(T,B)$ and we obtain region (I)where  $F_0(T,0)$ will dominate the low temperature thermodynamic properties. Similarly, the region (II) can be identified for $\frac{\omega_c}{\omega_0}\frac{\omega_{th}}{\omega_0}\sim 1$, where all the three terms of free energy are important in determining the low temperature thermodynamic properties. Lastly, the region (III) can be obtained for $\frac{\omega_c}{\omega_0}\frac{\omega_{th}}{\omega_0}>>1$ (we set it to 3.5), where the low temperature thermodynamic behaviour is determined by $\Delta_1F_0(T,B)$ or $\Delta_2F_0(T,B)$. Here, we should mention that we have differentiated two regimes if two terms ($F_0(T,0)$ and $\Delta_1F_0(T,B)$) of free energies differ by a order of magnitude.  These three regimes are clearly shown in the schematic phase diagram 1. We also tabulated some typical parameter values (B,T,and a) for a trapped Beryllium atom in contact with engineered phase reservoir in Table I. It is to be mentioned that we require nanostructures or microstructures to confine the charged particles and identify the three regimes by tuning external magnetic field B.\\
\indent
   Let us compare our results with that of standard coordinate-coordinate coupling case. Recently, we have shown that magnetic field dependence completely disappears ($\lim_{\omega\rightarrow 0}I_2 \simeq 0$ or $\Delta_1F_0(T,B)=0$ ) from the low-temperature thermodynamic properties, irrespective of $\mu(\omega)$, i.e., the nature of heat bath \cite{malay1}. So, there is no option to control thermodynamic properties at low temperatures for standard coordinate-coordinate coupling by varying the external parameters like, external magnetic field B or the confining length $a$. Thus, the thermodynamic properties at low temperatures are always determined by $F_0(T,0)$ for coordinate-coordinate coupling case. \\
\indent
Now, it is time to discuss about the effect of environment in determining low temperature thermodynamic properties. Suppose we are in regime (I) where the low temperature thermodynamic properties are determined by $F_0(T,0)$ alone and the entropy $S= -\frac{\partial F_0(T,B)}{\partial T}$ approaches zero as $T\rightarrow 0$ in conformity with third law of thermodynamics with a power law $S\sim T^{\delta}$ ($\delta=\nu+1$). Thus, the entropy falls off linearly to zero for the Ohmic bath ($\nu=1$). On the other hand, entropy vanishes to zero with a power $\delta>2$ and $1<\delta<2$ for the super-Ohmic and sub-Ohmic environment, respectively. If we move to regime (III) where thermodynamics is determined by $\Delta_1F_0(T,B)$, the entropy falls off to zero as $T\rightarrow0$ with a power law $T^{\beta}$ with $\beta=\nu+3$. Again, we find $3<\beta<4$, $\beta=4$ and $\beta>4$ for the sub-Ohmic, Ohmic and super-Ohmic cases respectively. Thus, we can say that the power of the power law depends on the nature of heat bath. Since, we can move from regime (I) to regime (III) for momentum dissipation by tuning $B$ and $a$, the power of the power law fall of entropy to zero as $T\rightarrow 0$ can also be tuned by varying B and a. It is also been observed that the entropy has a  faster decay ($\beta$) for momentum-momentum coupling compare to standard coordinate-coordinate coupling ($\delta$) for which we can only have regime (I).\\
\begin{table}
\begin{center}
    \begin{tabular}{|l|l|l|}
    \hline
    \multicolumn{3}{ |c| }{Arbitrary Heat Bath : Beryllium Ion} \\
    \hline
     Region I & Region II & Region III \\ \hline
     a=1$\mu$m & a= 1$\mu$m & a=1$\mu$m \\
     T=10 nK & T=10nK   & T=10nK \\
     B=100$\mu$T& B=10mT & B=100mT \\ \hline
     a=100nm & a= 100nm   & a=100nm \\
     T=1$\mu$K   & T=1$\mu$K & T=1 $\mu$K  \\
     B=10mT   & B=1T      & B=10T  \\ \hline
     \end{tabular}
\end{center}
\caption{Tabulated values of external parameters B, T, and a to observe three different regimes for a trapped Beryllium ion in contact with an engineered arbitrary heat reservoir.}
\end{table}
\indent
Let us consider the case without the confining potential, i.e., $\omega_0=0$. In this scenario, we have $\lim_{\omega\rightarrow 0}I_0=\frac{C(1-\nu)}{(A^2+C^2)}\omega^{1-\nu}$ , $\lim_{\omega\rightarrow 0}I_1=2\frac{C(1-\nu)}{(A^2+C^2)}\omega^{1-\nu}$, and $\lim_{\omega\rightarrow 0}I_2=0$. As a result we obtain the free energy as follows :
\begin{equation}
F_0(T,B)=-\hbar b(1-\nu)\Gamma(1-\nu)\zeta(2-\nu)\frac{m_r}{m\pi}\cos(\frac{\nu\pi}{2})\Big(\frac{\omega_{th}}{b}\Big)^{3-\nu}
\end{equation}
Thus, we can say that the effect of magnetic field disappears from low temperature thermodynamic properties for $\omega_0=0$, as it does not appear in free energy expression (18). This is just opposite to what happened for the coordinate-coordinate coupling. For coordinate-coordinate coupling, we have shown earlier in Ref. \cite{malay1,malay2} that the effect of magnetic field in the low temperature thermodynamic properties appears for $\omega_0=0$ case and the effect of $B$ disappears for $\omega_0\ne 0$.  On the other hand, entropy falls off to zero as $T\rightarrow 0$ with a power law $T^{\delta^{\prime}}$ with $\delta^{\prime}=2-\nu$ for $\omega_0=0$ with momentum dissipation. The power of the power law  is $\delta^{\prime}=1$ for Ohmic bath which is same as that of coordinate-cordinate coupling. Unlike the momentum dissipation, the prefactor of entropy, S(T), depends on the cyclotron frequency and the friction constant for coordinate-coordinate coupling.\\
\begin{figure}[h]
\begin{center}
{\rotatebox{0}{\resizebox{6cm}{5cm}{\includegraphics{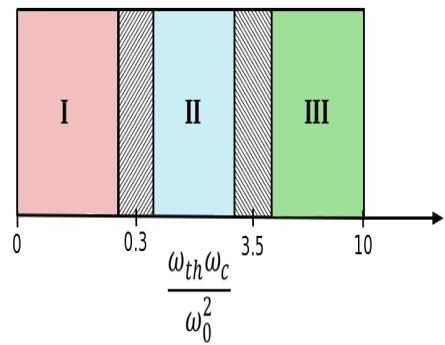}}}}
\caption{(color online) Approximate schematic sketch of different accessible regimes which can be obtained by varying the ratios of $\frac{\omega_c}{\omega_0}$ and $\frac{\omega_{th}}{\omega_0}$ for a trapped Beryllium ion in contact with an arbitrary heat bath. }
\end{center}
\end{figure}
Now, it is time to tell about the implementation of our control mechanism in physically realizable system.
\begin{figure}[h]
\begin{center}
{\rotatebox{0}{\resizebox{4cm}{4cm}{\includegraphics{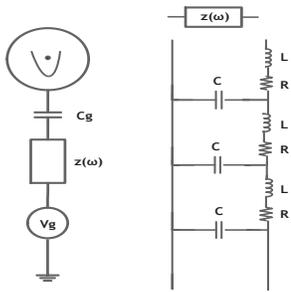}}}}
\caption{(color online) A schematic sketch of a physically realizable system with a trapped ion in contact with a Ohmic or sub-Ohmic environment }
\end{center}
\end{figure}
The possibility of controlling both the environment and the system-environment coupling open the doorway of controlling the low temperature thermodynamic behaviour at nanoscale. During the last few decades, the huge advancement in the field of laser cooling and trapping experimental techniques have made the way to confine a single ion in harmonic well at very low temperatures where quantum effects are predominant. For this purpose one can use a miniature version of the linear Paul trap \cite{raizen,jefferts}. A single laser cooled ion is theoretically equivalent to a charged particle moving in a harmonic well. Thus, it is now possible to arrange quantum Brownian motion in the context of trapped ions with the help of engineered reservoir. The advancement in reservoir engineering techniques \cite{myatt} pave the way to construct possible experiments aimed at simulating paradigmatic models of open quantum systems as the one considered in this paper. It is not only possible to construct "artificial" reservoir but also one can manipulate its spectral density and the coupling with the system \cite{myatt}. The possible way to implement a QBM model for an Ohmic and a sub-Ohmic environment has been discussed in Refs. \cite{b,tong,grob}. The same method can be extended straightforwardly to realize the Ohmic and sub-Ohmic environments considered here for a trapped Beryllium ion. In Refs. \cite{b,tong}, the cases of Ohmic and sub-Ohmic environment are modelled by an infinite RLC transmission line. As discussed in Ref. \cite{b} (page 63), the transmission line can be thought of made of discrete building blocks which consists of inductor (L) and resistor (R) are in series along one stringboard of the ladder and the capacitor (C) is on the horizontal support of the ladder as shown in figure 2. Thus, the impedance of an infinitely long transmission line is given by $Z(\omega)=\sqrt{\frac{R+i\omega L}{i\omega C}}$. Thus, it is evident that the Ohmic and sub-Ohmic environment can be realized from the LC dominant and R-dominant limit of the RLC transmission line, respectively \cite{b}. These would allow one to test in a controlled way a fundamental and ubiquitous model such as considered in this paper through  Quantum Brownian motion (QBM). In this respect, we should mention that QBM model with single trapped ion connected with Ohmic or non-Ohmic reservoir is simulated by Maniscalco {\em et al}\cite{manis1,manis2}. Experiments with single trapped ions have demonstrated the ability to engineer artificial environments and to control the relevant system-environment parameters \cite{myatt}. In our case, the trapped ion is a single $Be^{+}_9$ ion which is stored in a rf Paul trap \cite{jefferts} with a oscillation frequency of $\frac{\omega_0}{2\pi}= 11.2$ MHz . Then, the ion  can be laser cooled using sideband cooling with stimulated Raman transitions  between the $2^S_{1/2}$ (F = 2, $m_F =-2$) and  $2^S_{1/2}$ (F = 1, $m_F = -1)$ hyperfine ground states, which are denoted by "up" and "down", respectively. These states are separated by approximately 1.25 GHz. Reference \cite{myatt} represents the recent advancement to show how to couple a properly engineered reservoir with a quantum charged oscillator, e.g., the quantized center of mass motion of the trapped $Be^{+}_9$ ion. This trapped ion can be capacitively coupled with the impendence $Z(\omega)$. A schematic diagram of such an experimentally realizable system is drawn in figure 2. This situation is somewhat similar to the momentum dissipation case discussed in the context of quantum electrodynamic fluctuations of the macroscopic Josephson phase by H. Kohler {\em et al.} \cite{kohler}.
\subsection{Blackbody radiation bath}
In this case, the associated memory function is given by
\begin{equation}
\mu_d(\omega)=\frac{2q^2\Omega^2\omega}{3c^3(\omega+i\Omega)},
\end{equation}
where $\Omega$ is a cutoff frequency. It has been shown that in the large cut-off limit the memory function and the response function in the absence of magnetic field are given by \cite{ford1,ford2}:
\begin{eqnarray}
\mu_d(\omega)= -i\frac{M\omega}{1-i\omega\tau_e}, \nonumber \\
\alpha^{(0)}(\omega)= \Big\lbrack -\frac{M\omega^2}{1-i\omega\tau_e}+M\omega_0^2\Big\rbrack^{-1},
\end{eqnarray}
where, $M=m_r+\frac{2q^2\Omega}{3c^3}$ and $\tau_e=\frac{2q^2}{3Mc^3}$.
As a result, we have at low temperatures, i.e., only considering low frequencies contribution in (12) :
\begin{eqnarray}
\lim_{\omega\rightarrow 0}I_0=\frac{3m_r\tau_e}{M\omega_0^2}\omega^2, \nonumber \\
\lim_{\omega\rightarrow 0}I_1=-2\frac{m_r}{M}\frac{\omega_c^2\tau_e}{\omega_0^4}\omega^2, \nonumber \\
\lim_{\omega\rightarrow 0}I_2=8\frac{m_r}{M}\frac{\omega_c^2\tau_e}{\omega_0^4}\omega^2
\end{eqnarray}
Again using the result (15), we can obtain :
\begin{eqnarray}
F_0(T,0)=-\frac{3\pi^2\omega_0\tau_e}{5}\frac{m_r}{M}\Big(\frac{\omega_{th}}{\omega_0}\Big)^4\hbar\omega_0, \nonumber \\
\Delta F_0(T,B)=-\frac{2\pi^3}{9}\frac{m_r}{M}\omega_c\tau_e\Big(\frac{\omega_{th}}{\omega_0}\Big)^4\hbar\omega_c,
\end{eqnarray}
with $\Delta F_0(T,B)= \Delta_1 F_0(T,B) + \Delta_2 F_0(T,B)$.
\begin{table}
\begin{center}
    \begin{tabular}{|l|l|l|l|}
    \hline
    \multicolumn{3}{ |c| }{Radiation Heat Bath : Calcium Ion} \\
    \hline
     Region I &Region II   & Region III \\ \hline
     a=1$\mu$m & a= 1$\mu$m   & a=1$\mu$m \\
     T=10nK   & T=10nK     & T=10nK \\
     B=10$\mu$T   & B=1mT      & B=10mT \\ \hline
     a=100nm& a=100nm   & a=100nm \\
     T=1$\mu$K & T=1$\mu$K & T=1$\mu$K  \\
     B=1mT & B= 0.1T    & B=1T  \\ \hline
    \end{tabular}
\end{center}
\caption{Tabulated values of different controllable parameters like B, T, and a to observe different regimes for a trapped Calcium ion  in contact with an engineered radiation heat bath. }
\end{table}
Now, we can find the ratios of these two terms of the free energy :
\begin{eqnarray}
\frac{F_0(T,0)}{\Delta F_0(T,B)}=\frac{27}{10\pi}\Big(\frac{\omega_0}{\omega_c}\Big)^2.
\end{eqnarray}
One can again say that the power of the power law behaviour of different thermodynamic quantities can be controlled by varying  external parameters  magnetic field (B), and confining length (a) associated with $\omega_c$, and $\omega_0$, respectively. One can easily observe from the phase diagram (Fig 3) that it contains three regimes just like the arbitrary bath. Unlike the arbitrary bath, the three regimes for the radiation bath can be explored by varying only the ratio $\frac{\omega_c}{\omega_0}$ alone.  Let us consider the case of without confining potential, i.e., $\omega_0=0$ case. For this situation ($\omega_0=0$) we have :
\begin{eqnarray}
\lim_{\omega\rightarrow 0}I_0=-\tau_e, \nonumber \\
\lim_{\omega\rightarrow 0}I_1=0, \nonumber \\
\lim_{\omega\rightarrow 0}I_2=4\tau_e.
\end{eqnarray}
As a result, we have the free energy
\begin{equation}
F_0(T,B)=-\frac{\pi}{6}\hbar\omega_e\Big(\frac{\omega_{th}}{\omega_e}\Big)^2,
\end{equation}
where, $\omega_e= \tau_e^{-1}$. Unlike the coordinate-coordinate coupling, the free energy at low temperatures for the radiation heat bath with momentum dissipation is free from magnetic field \cite{malay1,malay2}. Also, we have observed entropy $s(T)= \frac{\pi k_B^2T}{3\omega_e\hbar}$ and entropy vanishes as $T\rightarrow 0$ but the temperature dependence and the prefactors are different from that of coordinate-coordinate case \cite{malay1,malay2}. Typical values of externally controlable parameters ($B$, $T$ and $a$) for the radiation heat bath are tabulated for the trapped Calcium ion in table II.  Once again we can observe three different regimes in the nanostructures or microstructures.
\begin{figure}[h]
\begin{center}
{\rotatebox{0}{\resizebox{6cm}{6cm}{\includegraphics{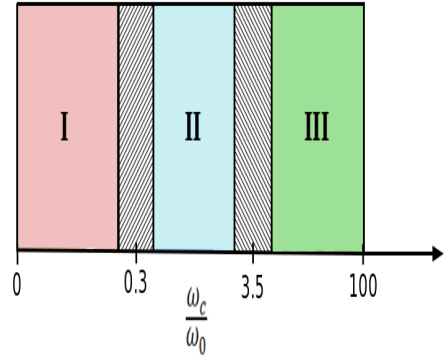}}}}
\caption{(color online) Different regimes for the trapped Calcium ion in contact with a radiation heat bath are plotted as a function of the ratio of $\frac{\omega_c}{\omega_0}$ alone.}
\end{center}
\end{figure}
Now, the question left, how one can mimic this radiation bath reservoir in the laboratory? This is achievable through the method discussed in Refs. \cite{moussa,zollar} in the context of a control of a cavity field state through an atom-driven field interaction. But, recently H. G. Barros et al \cite{barros} have reported on the realization of an efficient single-photon source using a single calcium ion trapped within a high-finesse optical cavity. This system shares some features with that of our case of single ion interacting with a radiation bath. A detailed description of the experimental setup can be found in \cite{russo}. In short, one can trap a single $Ca^{+}_{40}$ ion in a linear Paul trap situated in the center of a high-finesse optical cavity. The 2 cm long cavity  has asymmetric mirror reﬂectivities. The interaction between the trapped ion and the cavity field occurs via the atomic transition $|P_{1/2}, m = +1/2>  \leftrightarrow |D_{3/2}, m = −1/2>$, at 866 nm (cavity finesse of 70 000) with a maximum single-photon Rabi frequency of $2g_0 = 2\pi 3.2 MHz$. The photons which leave the cavity  are guided by a multimode fiber to a Hanbury Brown–Twiss (HBT) setup. So, one can easily realize the model study of a charged oscillator interacting with a radiation heat bath with the help of above mentioned setup.\\
Now, we  discuss about the effect of negative renormalized mass due to momentum dissipation on different thermodynamic quantities at low temperatures. We showed in our earlier publications that the effective mass is reduced as we increase the momentum coupling and the reduced mass is given by $m_r = \frac{m}{1+\sum_j\frac{g_j^2m}{m_j}}$ \cite{j,k}. This finding is in conformity with the observation of Cuccoli et al \cite{g} and Ankerhold et al \cite{h}. One should note that the renormalized mass arises from the $\omega^2$ term associated with the inertial term of the susceptibility expression. The effect of reduction of mass due to increase of momentum coupling can distinctly be observed in the low temperature thermodynamic properties. For $\omega_0\ne 0$, the quantum contribution to different thermodynamic quantities for the arbitrary heat bath case can be increased by increasing the strength of momentum-momentum coupling ($g_j$), as $m_r$ appears in the denominator of Eq. (16). On the other hand, as we increase $g_j$, the quantum contribution to different thermodynamic quantities reduces for the case of without the confining potential,i.e., $\omega_0=0$. For the radiation heat bath, the effect of reduction of effective mass (as we increase $g_j$) on different low temperature thermodynamic quantities is cancelled out due to appearance of the ratio $\frac{m_r}{M}$ in Eqs. (22). In this context we should mention that the negative renormalized mass is also discussed for several cases with position-position coupling \cite{ingold1,ingold2}. However, a difference between normal dissipation and momentum dissipation is indeed the appearence of renormalized mass in the potential term of the quantum Langevin equation derived from the Hamiltonian (1) with momentum dissipation \cite{j}. Although, this effect can be thought of as a secondary effect as the sign of the renormalized mass is usually read off from the inertial mass.

\section{Conclusions}
In this paper, we discuss the low temperature thermodynamic properties of a charged oscillator in the presence of an external magnetic field and is coupled with a quantum heat bath through momentum-momentum variables. Although, the validity of the third law is confirmed for different heat bath, but the power of the power law fall of the entropy as $T\rightarrow 0$ can be controlled by  external parameters : $B$ and $a$. Depending on the power of the power law, we can identify  different regimes for the arbitrary heat bath and radiation heat bath. Typical values of  external parameters ($B$ and $a$) to observe different regimes for a trapped Beryllium ion and trapped calcium ion in contact with different engineered reservoir are tabulated. Also, the effect of reduction of effective mass as we increase the momentum-momentum coupling strength $g_j$ on different QTF are discussed in details. In this context we have described a possible experimental realization of our control mechanism for the quantum thermodynamics for a trapped Beryllium ion interacting with Ohmic, sub-Ohmic or super-Ohmic engineered reservoir at nanoscale. On the other hand, we have described the experimental realization of engineered radiation bath in the context of trapped calcium ion.\\
\indent
 Now, with the advent of technological advancement reaching into the nano and quantum regime, and in view of the fundamentally different rules of quantum mechanics, there is utmost requirement to understand thermodynamics at the microscopic and nano scale where thermal fluctuations compete with quantum fluctuations. In that perspective our research will be helpful in controlling thermodynamic properties as well as understanding thermodynamics at micro and nano scale. We can conclude that our present study is relevant in the process of understanding thermodynamics at nano-scale as well as making of small scale thermal machines in which working fluid is a single trapped ion.\\
 \begin{acknowledgments}
MB acknowledge the financial support of IIT Bhubaneswar through seed money project SP0045.
\end{acknowledgments}


\begin{thebibliography}{99}
\bibitem{a}H.-P. Breuer and F. Petruccione, {\em The Theory of Open Quantum Systems} (Oxford University Press, Oxford, 2002).
\bibitem{b}U. Weiss, {\em Quantum Dissipative Systems} (World Scientific, Singapore, 2008).
\bibitem{c}A. O. Caldeira and A. J. Leggett, Ann. of Phys. {\em 149}, 374 (1983).
\bibitem{d}P. Ullersma, Physica {\em 32}, 90 (1966).
\bibitem{e}G. W. Ford, J. T. Lewis, and R. F. O'Connell, Phys. Rev. A {\em 37}, 4419 (1988).
\bibitem{f}A. J. Leggett, Phys. Rev. B {\em 30}, 1208 (1984).
\bibitem{g}A. Cuccoli, A. Fubini, V. Tognetti, and R. Vaia, Phys. Rev. E {\em 64}, 066124 (2001).
\bibitem{h}J. Ankerhold, and E. Pollak, Phys. Rev. E {\em 75}, 041103 (2007).
\bibitem{i}F. Sols and I. Zapata, in New Developments on Fundamental Problems in Quantum Physics, eds. M. Ferrero and A. van der Merwe (Kluwer, Dordrecht,1997).
\bibitem{makhno}Y. A. Makhnovskii and E. Pollak, Phys. Rev. E {\em 73}, 041105 (2006).
\bibitem{fermi}E. Fermi, Phys. Rev. {\em 75}, 1169 (1949).
\bibitem{sturrock}P. A. Sturrock, Phys. Rev. {\em 141}, 186 (1966).
\bibitem{cole}D. C. Cole, Phys. Rev. E {\em 51}, 1663 (1995).
\bibitem{ingold1}Benjamin Spreng, Gert-Ludwig Ingold, Ulrich Weiss, Eur. Phys. Lett. {\em 103}, 60007 (2013).
\bibitem{ingold2}Robert Adamietz, Gert-Ludwig Ingold, Ulrich Weiss, Eur. Phys. J. B {\em 87}, 90 (2014).
\bibitem{sols1}H. Kohler, and F. Sols, New J. Phys. {\em 8}, 149 (2006).
\bibitem{sols2}H. Kohler, and F. Sols, Physica A {\em 392}, 1989 (2013).
\bibitem{j}S. Gupta, and M. Bandyopadhyay, Phys. Rev. E {\em 84}, 041133 (2011).
\bibitem{k}S. Gupta, and M. Bandyopadhyay, J. Stat. Mech. : Theory and Expt. P04034 (2013).
\bibitem{ingold3}Gert-Ludwig Ingold, Peter H{\"a}nggi, Peter Talkner, Phys. Rev. E {\em 79}, 061105 (2009).
\bibitem{ingold4}Peter H{\"a}nggi, Gert-Ludwig Ingold and Peter Talkner, New J. Phys. {\em 10}, 115008 (2008).
\bibitem{ingold5}P. H{\"a}nggi and G.-L. Ingold, Acta Phys. Pol. B {\em 37}, 1537 (2006).
\bibitem{l}J. H. Van Vleck, The theory of Electric and Magnetic Susceptibilities (London, Oxford University Press, 1932).
\bibitem{m}R. B. Laughlin, Phys. Rev. B {\em 23}, 5632 (1981).
\bibitem{n}V. L. Ginzburg, and D. A. Kirzhntis, High Temperature Superconductivity (New York, Consultants Bureau, 1982).
\bibitem{li}X. L. Li, G. W. Ford, and R. F. O`Connell, Phys. Rev. A {\em 42}, 4519 (1990).
\bibitem{wang}C.-Y. Wang, J.-D. Bao, Chin. Phys. Lett. {\em 25}, 429 (2008).
\bibitem{rmp}A. J. Leggett, S. Chakravarty, A. T. Dorsey, M. P. A. Fisher, A. Garg, and W. Zwerger, Rev. Mod. Phys. {\em 59}, 1 (1987).
\bibitem{o}A. Shnirman et al., Phys. Scr. {\em T102}, 147 (2002).
\bibitem{malay1}M. Bandyopadhyay, J. Stat. Phys. {\em 140}, 603 (2010);J. Stat. Mech. : Theory \& Expt. P05002 (2009).
\bibitem{malay2}M. Bandyopadhyay, and S. Dattagupta, Phys. Rev. E {\em 81}, 042102 (2010).
\bibitem{raizen}M.G. Raizen, J.M. Gilligan, J.C. Bergquist, W.M. Itano, and D.J. Wineland, Phys. Rev. A {\em 45}, 6493 (1992).
\bibitem{jefferts}S.R. Jefferts, C. Monroe, E.W. Bell, and D.J. Wineland, Phys. Rev. A  {\em 51}, 3112 (1995).
\bibitem{myatt}Q. A. Turchette, C. J. Myatt, B. E. King, C. A. Sackett, D. Kielpinski, W. M. Itano, C. Monroe, and D. J. Wineland, Phys. Rev. A {\em 62}, 053807 (2000); C. J. Myatt et al., Nature (London) {\em 403}, 269 (2000).
\bibitem{tong}N.-H. Tong and M. Vojta, Phys. Rev. Lett. {\em 97}, 016802 (2006).
\bibitem{grob}S. Gr{\"o}blacher, A. Trubarov, N. Prigge, M. Aspelmeyer, and J. Eisert, arXiv:1305.6942v1 (2013).
\bibitem{manis1}S. Maniscalco, J. Piilo, F. Intravaia, F. Petruccione, and A. Messina, Phys. Rev. A {\em 69}, 052101 (2004).
\bibitem{manis2}J. Piilo and S. Maniscalco, Phys. Rev. A {\em 74}, 032303 (2006).
\bibitem{poya}J. F. Poyatos, J. I. Cirac, and P. Zoller, Phys. Rev. Lett. {\em 77}, 4728 (1996).
\bibitem{kohler}H. Kohler, F. Guinea, and F. Sols, Annals of Phys. {\em 310}, 127 (2004).
\bibitem{ford1}G. W. Ford, and R. F. O'Connell, Physica E {\em 29}, 82, (2005).
\bibitem{ford2}G. W. Ford, J. T. Lewis, and R. F. O'Connell, Phys. Rev. A {\em 36}, 1466 (1987).
\bibitem{moussa}C. J. Villas-Boas, F. R. de Paula, R. M. Serra, and M. H. Y. Moussa, Phys. Rev. A {\em 68}, 053808
( 2003)
\bibitem{zollar}N. Lutkenhaus, J. I. Cirac, and P. Zoller, Phys. Rev. A {\em 57}, 548 (1998)
\bibitem{barros} H. G. Barros {\em et al.} New J. Phys. {\em 11}, 103004 (2009)
\bibitem{russo}C. Russo  {\em et al.}  Appl. Phys. B {\em 95}, 205 (2009)
\end{thebibliography}
\end{document}